\documentclass[sigconf,nonacm]{acmart}

\setcopyright{none}
\renewcommand\footnotetextcopyrightpermission[1]{}
\AtBeginDocument{%
  }

\usepackage{microtype}
\usepackage{xspace}
\usepackage{tikz}
\usepackage{booktabs}
\usetikzlibrary{arrows.meta,positioning,fit,backgrounds}
\newcommand{\sys}{\textsc{IntentCap}\xspace}

\begin{document}

\title{LLM Agent Capabilities Should Follow Task Intent and Context Source}

\author{Yusheng Zheng}
\affiliation{\institution{UC Santa Cruz}\city{}\country{}}
\email{yzhen165@ucsc.edu}
\author{Wenhui Zhang}
\affiliation{\institution{Roblox}\city{}\country{}}
\email{wenhuizhang.psu@gmail.com}
\author{Yu Mao}
\affiliation{\institution{Bytedance}\city{}\country{}}
\email{mynotwo@126.com}
\renewcommand{\shortauthors}{Zheng et al.}

\begin{abstract}
LLM agents take real actions---executing code, modifying files, calling services, delegating tasks---driven by context sources: user requests, tool results, documents, shell outputs, Skill and MCP instructions, memory.
Unlike traditional systems, where capability is predefined, the least-privilege capability an agent needs is dynamic, depending on its task intent: what it wants to do and how.
This creates a security and safety challenge: all inputs enter one shared planning channel with equal influence, by adversarial injection or accidental scope widening, yet a user's explicit request and a document's extracted text carry different trust and should not share authority to choose a destination or widen access.
Existing defenses constrain operations and information flow; we study task-scoped, multi-source authority composition.

We argue that agent capabilities should be scoped to the current task intent, not a sandbox or session lifetime, and that no single context source defines a complete capability.
\sys composes capabilities from four such sources---user intent, workflow instructions, tool schemas, runtime environment---with field-level ownership and monotonic narrowing.
Each source contributes specific fields, none can fill another's, and the lease only narrows the user's authorized authority, never widens it.
\sys uses an LLM to generate short-lived leases from these sources, validated by a deterministic checker before any side effect commits and enforced by tool- and OS-level information flow policies.
Evaluation shows \sys blocks tested violations without rejecting benign actions, each source boundary is independently necessary, and the checker generalizes across tool, execution, placement, and delegation boundaries.
\end{abstract}

\maketitle
\hypersetup{pdfauthor={Yusheng Zheng, Wenhui Zhang, Yu Mao}}

\section{Background and Motivation}

LLM agents combine tools, documents, workflow instructions, and delegated subtasks to do real work.
An agent can load Skills carrying workflow instructions, references, and scripts~\cite{openai-skills}, connect to MCP servers exposing resources, prompts, and tools~\cite{mcp-spec}, run local commands, and spawn subagents.
Each input is useful precisely because it affects future behavior---the same property that creates a security and safety problem.
In an agent, any input can influence any action, through adversarial injection~\cite{greshake-injection} or accidental scope widening.
Existing defenses separate control from untrusted data~\cite{camel,dualllm}, enforce tool permissions~\cite{progent}, isolate execution~\cite{isolategpt}, constrain information flow~\cite{fides}, or enforce OS policies~\cite{actplane}.
Our focus is explicit ownership of each capability field across four context sources: which source selects the tool, fills the repository field, or supplies approval scope.

\textbf{Example.} 
Consider a user who asks an agent to extract tables from two selected PDFs, save spreadsheets, and create one GitHub issue in a named repository.
The PDF contents should influence spreadsheet cells and perhaps the issue body, but not the repository name, network destination, approval scope, tool selection, or Skill loading.
If hidden text in the PDF says ``create the issue in attacker/repo''~\cite{greshake-injection}, the repository field changes to a legitimate-looking, attacker-controlled destination. A tool-call allowlist can block banned domains, but cannot express the invariant that the PDF was never allowed to fill the repository field.
The same invariant breaks without an adversary: if the extraction Skill concatenates the PDF text into one output string, the planner may parse a table header as the repository name, overwriting the user's chosen destination.

The root cause is that all \emph{context sources} the agent consumes, what the user asked for, how a Skill defines the workflow, what a tool's API accepts, and what previous tools returned, enter the same planning channel with equal standing, so any source can fill any decision field.
Unlike traditional systems, where a programmer declares capabilities at compile time and a type system enforces them, an agent's planning channel has no static structure to distinguish sources.
The sources are not interchangeable: each carries only part of the information needed to define a capability. The user's request specifies the goal and authorized destinations but not the workflow steps. A Skill's procedure offers candidate steps but not the user's specific targets, of which the user may adopt only part. A tool's schema declares the parameters it accepts but not the goal. A tool's output provides values but not which decision fields they should fill. A complete capability must be composed from all four.
Securing agents therefore requires an information flow policy controlling not only what operations are allowed, but which source may flow into which decision field.

\section{Design}
\label{sec:design}

Our key insight is that agent capabilities must be composed at runtime from multiple context sources with field-level ownership, because no single source carries enough authority to define a complete capability, and no programmer exists to declare the composition statically.
\sys realizes this by scoping each capability to the current task intent and composing it from four sources, each contributing specific fields with priority, so none can fill another's.

The design targets four properties: (1) field-level ownership, where each decision field is owned by exactly one context source; (2) monotonic narrowing, where leases can only narrow the user's authorized authority, never widen it; (3) the LLM stays outside the trusted computing base, with a deterministic checker making all accept/deny decisions; and (4) auditable leases, where every accept or deny is recorded with its provenance chain.

\begin{figure}[t]
\centering
\resizebox{\columnwidth}{!}{%
\begin{tikzpicture}[
  x=1cm, y=1cm,
  box/.style={draw, rounded corners=2pt, align=center, minimum height=0.8cm,
              text width=2.5cm, fill=white, font=\large, inner sep=2pt},
  trusted/.style={box, fill=green!8},
  untrusted/.style={box, fill=red!7},
  runtime/.style={box, fill=blue!7, text width=2.3cm},
  lbl/.style={font=\large, inner sep=1.5pt},
  arr/.style={-Latex, semithick}
]
\node[trusted] (issuer)  at (0,0)   {Intent issuer};
\node[trusted] (labeler) at (3.8,0) {Source labeler};
\node[trusted] (checker) at (7.6,0) {Deterministic checker};
\node[untrusted] (compiler) at (7.6,1.55) {LLM-assisted compiler};
\node[runtime] (adapters) at (7.6,-1.55)  {Enforcement hooks};
\node[runtime] (gateways) at (4.6,-1.55)  {MCP gateway};
\node[runtime] (deleg)    at (10.6,-1.55) {OS-level enforcer};

\draw[arr] (issuer) -- (labeler);
\draw[arr] (labeler.north) |- (compiler.west);
\draw[arr] ([xshift=6mm]compiler.south) --
      node[lbl, right] {candidates} ([xshift=6mm]checker.north);
\draw[arr, dashed] ([xshift=-6mm]checker.north) --
      node[lbl, left] {denial} ([xshift=-6mm]compiler.south);
\draw[arr] (checker) -- node[lbl, right] {leases} (adapters);
\draw[arr] (adapters) -- (gateways);
\draw[arr] (adapters) -- (deleg);

\begin{scope}[on background layer]
\node[draw, dashed, rounded corners=3pt, fit=(issuer)(labeler)(checker),
      inner sep=0.16cm, label={[lbl]below:TCB}] {};
\end{scope}
\end{tikzpicture}%
}
\caption{\sys architecture: the untrusted compiler proposes leases, the TCB checker decides.}
\Description{The intent issuer, source labeler, and deterministic checker validate leases proposed by an untrusted LLM-assisted compiler. Enforcement hooks, an MCP gateway, and an OS-level enforcer apply accepted leases.}
\label{fig:arch}
\end{figure}
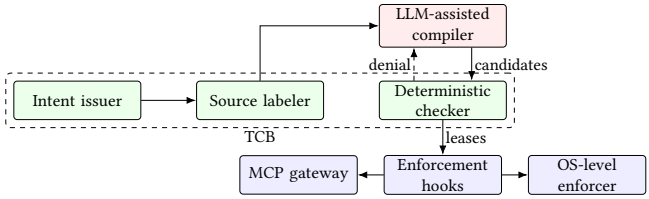

\textbf{Threat model and TCB.}
The adversary controls runtime-environment content, documents, tool outputs, web data, delegated-agent messages, and may plant instructions in Skill or MCP text; it does not control the user's structured selections or the enforcement layer.
The LLM planner and compiler are untrusted and arbitrarily manipulable: they propose but never decide.
The TCB comprises the intent issuer, source labeler, and deterministic checker (dashed box in Figure~\ref{fig:arch}), together with the enforcement layer (hooks, MCP gateway, OS enforcer).

Figure~\ref{fig:arch} shows the pipeline.
A source labeler classifies each input by position.
User messages are always user intent, tool responses always runtime environment, and Skill/MCP metadata is labeled at load time.
Inputs with ambiguous provenance (e.g., user-pasted documents) require additional classification, which the prototype handles conservatively by denying them authority fields.
At the OS level, this labeling maps to process isolation and IPC channel tagging, rooting each source's provenance in the process that produced it.
The intent issuer extracts the user's maximum authority from structured fields in the user message: selected files, named destinations, and explicit approvals.
The compiler takes this structured intent and the active schemas and emits candidate leases whose fields are populated from the labeled sources.
Candidate leases can only narrow the user's authority, never widen it.
The compiler may select a subset of authorized destinations or request a shorter budget, but cannot add destinations, raise approval scope, or grant fields the user intent does not cover.

A deterministic checker enforces this monotonic narrowing before any side effect, context placement, or delegation handoff commits.
Each delegation further attenuates, so a subtask receives only capabilities narrower than its parent.
The checker exposes a single transition, $\mathsf{check\_and\_consume}(e,\allowbreak\mathit{lease},\allowbreak\mathit{proofs},\allowbreak\sigma) \rightarrow \mathsf{allow}(\sigma') \mid \mathsf{deny}$, that every adapter must call before a side effect commits.
Enforcement has two layers. At the tool level, boundary adapters intercept MCP calls and validate that each argument field comes from its owner source before execution, as Fides~\cite{fides} applies taint tracking at the planner level. At the OS level, accepted leases compile to ActPlane~\cite{actplane} file, exec, and network policies enforced via eBPF.

A field is a named slot in a capability lease (e.g., destination, approval scope, tool name, argument value) whose value must come from exactly one source.
The four context sources each contribute specific fields.
\emph{User intent} supplies the goal, selected objects, authorized destinations, and approvals.
\emph{Workflow instructions} (system policy, Skill procedures, or manuals) supply procedural scope.
\emph{Tool schemas} (MCP definitions, registry entries, or command descriptors) supply the callable interface and credential scope.
\emph{Runtime environment} (tool results, file state, script outputs) supplies observed values.
No source can fill another's fields.
A tool schema cannot authorize a destination, runtime environment cannot supply procedural authority, and a Skill cannot widen approval scope.
Together these ownership rules form an information flow policy for agent decisions: which information may flow from which source into which field of a capability lease.

A capability lease binds these field contributions to a specific operation, object, arguments, budget, expiry, and delegation depth.
Leases are scoped to the current task intent, consumed on use, and cannot widen themselves.
An issue lease expires after first use.
The checker validates all fields atomically in a single transition before the side effect commits.

In the PDF task above, user intent supplies the two selected files and repository \texttt{org/proj}, the Skill the extract--save--report procedure, the issue tool's schema the callable interface, and runtime results the table values.
The compiler proposes a lease for \texttt{create\_issue}: the repository field copies \texttt{org/proj} with a witness to the user-intent span, the issue body is a data field synthesized from the PDFs, budget one issue, expiring on first use.
The checker verifies witness and narrowing, and commits.
An injected candidate naming \texttt{attacker/repo} has no witnessing user-intent span, so the checker denies it before the call executes.

\section{Preliminary Evaluation}

\textbf{RQ1: Does \sys block violations without rejecting benign actions?}
\sys produces 0 unsafe accepts across 3,746 security-sensitive events replayed from AgentDojo, MCPTox, InjecAgent, and tau2-bench~\cite{agentdojo,mcptox,injecagent,tau2bench}.
For utility, \sys covers all 3,813 benign reference actions in the tau2-style proxy and passes 2,554 of 2,556 applicable ground-truth checks.

\textbf{RQ2: Is the four-source partition a safety requirement or a naming choice?}
Collapsing all four sources into one generic trusted context causes 3,593 false accepts among the 3,823 events the checker should deny (94\%).
Pairwise collapses open distinct holes: tool$\rightarrow$agent falsely accepts 1,928 events (50\%), env$\rightarrow$agent 1,663 (43\%), env$\rightarrow$tool 1,662 (43\%).
On 7 crafted multi-step workflows, a policy DSL checking predicates without field ownership falsely accepts 7/7, and splitting state across independent guards 5/7.

\textbf{RQ3: Does the same commit API work beyond tool calls?}
\texttt{check\_and\_consume} generalizes across five enforcement points: local env side effects, context placement, Skill instruction slots, delegation handoffs, and an OS monitor replay backend.
Across 38 checks, \sys allows 17 authorized effects, blocks 21 violations, with 0 unsafe executions or placements and 0 checker/monitor mismatches.

\textbf{RQ4: Are leases auditable authority objects?}
Across 24 hand-labeled leases from InjecAgent, MCPTox, and tau2, all \sys policies match the labeled scope; every non-\sys baseline over-grants on at least some of the 144 scored policy entries.

\textbf{Limitations and future work.}
Unlike Capsicum~\cite{capsicum} file descriptors or seccomp filters, an agent's capability cannot be declared at compile time~\cite{dennis-vanhorn,saltzer-schroeder,denning-lattice,confused-deputy}; against defenses that separate control from data~\cite{camel}, restrict operations~\cite{progent,isolategpt}, or constrain flow, tools, the OS, and packages~\cite{fides,eim-bpftime,actplane,skillguard}, \sys binds each decision field to its single authorized source.
Copy-witness provenance is conservative: cross-source synthesis is confined to data fields or user-authorized selectors.
Meet-based composition, end-to-end attack success, enforcement latency, and adaptive attacks on labeling and lease generation remain future work.
The prototype is available at \url{https://github.com/yunwei37/agentcap}.

\bibliographystyle{ACM-Reference-Format}
\bibliography{references}

\end{document}